\documentclass[preprint]{aastex}
\usepackage{natbib, aas_macros}
\shorttitle{PAH properties in M17SW}
\shortauthors{Yamagishi, M. et al.}

\begin{document}

\title{Spatial variations of PAH properties in M17SW revealed by $Spitzer$/IRS spectral mapping}

\author{M. Yamagishi\altaffilmark{1}, H. Kaneda\altaffilmark{2}, D. Ishihara\altaffilmark{2}, S. Oyabu\altaffilmark{2}, T. Suzuki\altaffilmark{2}, T. Onaka\altaffilmark{3}, T. Nagayama\altaffilmark{4}, T. Umemoto\altaffilmark{5,6}, T. Minamidani\altaffilmark{5,6}, A. Nishimura\altaffilmark{2}, M. Matsuo\altaffilmark{4,5}, S. Fujita\altaffilmark{5,7}, Y. Tsuda\altaffilmark{8}, M. Kohno\altaffilmark{2}, and S. Ohashi\altaffilmark{3}}
\email{yamagish@ir.isas.jaxa.jp}

\altaffiltext{1}{Institute of Space and Astronautical Science, Japan Aerospace Exploration Agency, Chuo-ku, Sagamihara 252-5210, Japan}
\altaffiltext{2}{Graduate School of Science, Nagoya University, Furo-cho, Chikusa-ku, Nagoya 464-8602, Japan}
\altaffiltext{3}{Graduate School of Science, The University of Tokyo, 7-3-1 Hongo, Bunkyo-ku, Tokyo 113-0033, Japan}
\altaffiltext{4}{Graduate School of Science and Engineering, Kagoshima University, 1-21-35 Korimoto, Kagoshima 890-0065, Japan}
\altaffiltext{5}{Nobeyama Radio Observatory, National Astronomical Observatory of Japan (NAOJ), National Institutes of Natural Sciences (NINS), 462-2, Nobeyama, Minamimaki, Minamisaku, Nagano 384-1305, Japan}
\altaffiltext{6}{Department of Astronomical Science, School of Physical Science, SOKENDAI (The Graduate University for Advanced Studies), 2-21-1, Osawa, Mitaka, Tokyo 181-8588, Japan}
\altaffiltext{7}{Graduate School of Pure and Applied Sciences, University of Tsukuba, 1-1-1 Tennodai, Tsukuba, Ibaraki 305-8577, Japan}
\altaffiltext{8}{Graduate School of Science and Engineering, Meisei University, 2-1-1 Hodokubo, Hino, Tokyo 191-0042}

\begin{abstract}
We present $Spitzer$/IRS mid-infrared spectral maps of the Galactic star-forming region M17 as well as IRSF/SIRIUS Br$\gamma$ and Nobeyama 45-m/FOREST $^{13}$CO ($J$=1--0) maps.
The spectra show prominent features due to polycyclic aromatic hydrocarbons (PAHs) at wavelengths of 6.2, 7.7, 8.6, 11.3, 12.0, 12.7, 13.5, and 14.2~$\micron$.
We find that the PAH emission features are bright in the region between the HII region traced by Br$\gamma$ and the molecular cloud traced by $^{13}$CO, supporting that the PAH emission originates mostly from photo-dissociation regions.
Based on the spatially-resolved $Spitzer$/IRS maps, we examine spatial variations of the PAH properties in detail.
As a result, we find that the interband ratio of PAH 7.7~$\micron$/PAH 11.3~$\micron$ varies locally near M17SW, but rather independently of the distance from the OB stars in M17, suggesting that the degree of PAH ionization is mainly controlled by local conditions rather than the global UV environments determined by the OB stars in M17.
We also find that the interband ratios of the PAH 12.0~$\micron$, 12.7~$\micron$, 13.5~$\micron$, and 14.2~$\micron$ features to the PAH~11.3~$\micron$ feature are high near the M17 center, which suggests structural changes of PAHs through processing due to intense UV radiation, producing abundant edgy irregular PAHs near the M17 center.
\end{abstract}


\section{Introduction}

Spectral bands due to polycyclic aromatic hydrocarbon (PAH) emissions are dominant features in the near- and mid-infrared (IR; 3--20~$\micron$).
Since PAHs are excited by far-UV photons (6--13.6~eV), the PAH emission features are characteristic of photo-dissociation regions (PDRs).
The PAH emission features are observed at wavelengths of 3.3, 6.2, 7.7, 8.6, 11.3, 12.0, 12.7, 13.5, 14.2, 15.8, 16.4, 17.4, 17.8 and 18.9~$\micron$, which are attributed to vibrations of C-H or C-C bonds in hydrocarbons.
Those features, especially main features at 6.2, 7.7, 8.6, 11.3, and 12.7~$\micron$, are theoretically and observationally well-studied (e.g., \citealt{Chan01, Draine07, Tielens08, Bauschlicher09}).
Past studies have shown that the PAH interband ratios are useful probes to study the properties of PAHs.
Among these, the degree of PAH ionization is best studied through interband ratios involving the 6.2, 7.7 and 11.3~$\micron$ PAH interband ratios.
Here the 6.2 and 7.7~$\micron$ bands, due to C-C vibrations, are representative of ionized PAHs and the 11.3~$\micron$ band, due to C-H vibrations, representative of neutral PAHs (e.g., \citealt{Allamandola99, Peeters02}).
It is expected that the degree of PAH ionization is relatively high near exciting sources while that is low in molecular clouds.
Another important property is the edge structure of PAHs, which is probed by using the interband ratios of the PAH features at 11.3, 12.0, 12.7, 13.5, and 14.2~$\micron$; likely origins of these features are all C-H in-plane bending, but the numbers of adjacent C-H bonds in a benzene ring are different (PAH~11.3~$\micron$: solo; PAH~12.0~$\micron$: duo; PAH~12.7~$\micron$: trio; PAH~13.5~$\micron$ and PAH~14.2~$\micron$: quartet; \citealt{Draine03, Tielens05}).
For example, edgy PAHs are expected to show the strong PAH~12.0, 12.7, 13.5, and 14.2~$\micron$ features relative to the PAH~11.3~$\micron$ feature.
The structures of PAHs may change from region to region depending on the surrounding radiation field (\citealt{Boersma13, Kaneda14}).
Therefore examining the PAH ionization together with the PAH edge structure may be helpful to discuss variations of the PAH properties.

Recently, variations in the PAH properties have been intensively studied for a variety of targets mainly with $Spitzer$ and $AKARI$ (e.g., \citealt{Peeters02, Smith07a, Boersma12, Yamagishi12}).
Most of such studies, however, discussed PAHs in individual areas, and did not intensively examine spatial variations of PAHs.
In order to examine the effects of the surrounding interstellar environment on the PAH properties, spatially-resolved observations are essential (e.g., \citealt{Crete99, Rapacioli05, Berne07, Sakon07, Kaneda05, Kaneda08, Fleming10, Yamagishi10, Berne12, Pilleri12, Egusa13, Croiset16}).
One of the most intensive spatially-resolved studies in PAHs was carried out by \citet{Boersma13, Boersma14, Boersma15}.
They analyzed $Spitzer$/IRS spectral maps of the reflection nebula NGC~7023 and decomposed the observed spectra to the emission features from ionized and neutral PAHs.
As a result, they found clear spatial variations in the degree of PAH ionization along the direction from the exciting B-type star to PDR and molecular-cloud regions.
\citet{Stock16} also examined the PAH interband ratios of seven Galactic HII regions and three reflection nebulae with $Spitzer$/IRS spectral maps, although they did not discuss spatial variations of the ratios in the target regions.
\citet{Haraguchi12} examined the degree of PAH ionization in the Orion nebula based on their ground-based near-IR narrow-band observations.
The number of such spatially-resolved studies of PAHs is, however, still limited.

In this paper, we present $Spitzer$/IRS spectral maps of the mid-IR PAH features in the Galactic star-forming region M17 as well as Br$\gamma$ and $^{13}$CO($J$=1--0) maps covering the same region.
M17 is young ($\sim$1~Myr; \citealt{Hanson97}) and one of the well-studied active Galactic star-forming regions (e.g., \citealt{Stutzki90, Giard92, Giard94, Cesarsky96, Povich07}), which contains more than 100 OB stars in the central cluster, NGC~6618 (\citealt{Lada91}).
Among them, the most active ionizing source is CEN1, a binary of O4+O4 stars (\citealt{Chini80}).
Assuming the distance of 2~kpc (\citealt{Xu11}), the area of the spectral maps in the present study is $1.2\times1.2$~pc$^2$ which is 80 times larger than that in the study of NGC~7023 (\citealt{Boersma13}).
Based on the wide-area spectral maps, we examine the effects of the intense star-forming activity on the properties of PAHs in detail.

\section{Observations and Data Reduction}

\subsection{$Spitzer$/IRS}

We analyzed archival $Spitzer$/IRS spectral mapping data (AORKEY: 17976320, 17976576, 17977344, and 17977600), which were retrieved from the $Spitzer$ Heritage Archives.
The observed area is shown in Fig.~\ref{jiang_area}.
The used module is short-wavelength-low-resolution (SL) covering a wavelength range of 5.2--14.5~$\micron$ with a spectral resolution of $R$=64--128 (\citealt{Houck04}).
We used CUbe Builder for IRS Spectra Maps ver. 1.8 (CUBISM; \citealt{Smith07b}) with the default setting to reduce the data, and obtained spectral and its uncertainty cubes for the three orders named SL1 (7.4--14.5~$\micron$), SL2 (5.2--7.7~$\micron$), and SL3 (7.3--8.6~$\micron$), separately.
We checked bad pixels with eyes and removed them.
In the data reduction, background spectra were not considered.
The major background in the mid-IR is the zodiacal light, the spectra of which have smooth continua without prominent spectral features (\citealt{Reach96}).
The intensity of the zodiacal light is $\sim$60~MJy/sr at a wavelength of 18~$\micron$ around M17 (\citealt{Kondo16}).
The contribution of the zodiacal light in M17 spectra is negligible near the center, while that is not in outer regions.
We, therefore, do not discuss the continuum emission in the present study.
After unifying the pixel scale from the original size to 3.$\arcsec$6 for the three FITS cubes to correct differences in the spatial resolution at each wavelength range, the intensity levels for SL1 and SL2 were adjusted to coincide with SL3 in the overlapped wavelength ranges.
Finally, we obtained 990 independent mid-IR spectra.

\subsection{IRSF/SIRIUS}

We performed narrow-band imaging of M17 with the SIRIUS camera on the IRSF 1.4~m telescope (\citealt{Nagashima99, Nagayama03}).
SIRIUS has a field of view of 7.$\arcmin$7 $\times$ 7.$\arcmin$7 with a pixel scale of 0.$\arcsec$45.
We observed M17 using the two narrow-band filters tuned for the Pa$\beta$ 1.28~$\micron$ and Br$\gamma$ 2.16~$\micron$ lines, simultaneously.
The effective band widths of the filters are 0.029~$\micron$ for Pa$\beta$ and 0.038~$\micron$ for Br$\gamma$.
The observation was carried out on 2013 June 10 with an integration time of 75 seconds and 20 dithering.
The observed area is also shown in Fig.~\ref{jiang_area}.
In the present study, we analyzed only the Br$\gamma$ image which is less affected by the interstellar extinction than the Pa$\beta$ image.

We reduced the image data based on the standard data reduction procedure including dark subtraction, flat-fielding, sky subtraction and dithered-image-combining.
Since we did not subtract a continuum image, free-free continuum emission may be contaminated in the Br$\gamma$ image.
Photometric calibration was performed by comparing point-source fluxes in the image with those in the 2MASS Point Source Catalog (\citealt{Skrutskie06}), where we assumed that the magnitude of each point source is the same between the $K_s$-band and the narrow-band images. 
We used sufficiently isolated (11) point sources with the $K_s$-band flux in a range of 9.0--11.0~mag and with errors smaller than 0.05 mag.
As a result, the uncertainty of the photometric calibration coefficient is $\sim$5~\%.
The final 1$\sigma$ noise level is $7.0\times10^{-9}~\mathrm{W/m^2/sr}$.

\subsection{Nobeyama 45-m/FOREST}

We analyzed CO mapping data of M17, which were taken in the framework of the FUGIN (FOREST Ultra-Wide Galactic plane survey In Nobeyama; \citealt{Minamidani16a}) legacy survey covering a wide area of the Galactic plane ($l$=10--50$^\circ$, 198--236$^\circ$, $|b|<1^\circ$) with a high spatial resolution of 15$\arcsec$ and 1$\sigma$ noise level of 0.4~K in the $T_{\mathrm{mb}}$ scale at 1 km/s velocity resolution.
FOREST (\citealt{Minamidani16b}) is a four-beam receiver, and has a capability of simultaneous on-the-fly mapping observations using the three CO lines, $^{12}$CO ($J$=1-0), $^{13}$CO ($J$=1-0), and C$^{18}$O ($J$=1-0).
Pointing errors were corrected every 1.5 hours by observing SiO maser sources.
As a result, the pointing accuracy of the telescope is kept to be $<5\arcsec$.
We used calibrated $^{13}$CO map (internal team release version 1.3) because it is likely optically thin and traces overall molecular cloud structures in star forming regions.
The $^{13}$CO map has spatial resolution of 18$\arcsec$ and 1$\sigma$ noise level of $\sim$0.8~K in the $T_{\mathrm{mb}}$ scale at 1~km/s velocity resolution as an intermediate product of the survey project.
The detail of the survey observations and data reductions will be described in Umemoto et al. in prep.

\section{Analyses and Results}

\begin{figure}
\epsscale{.5}
\plotone{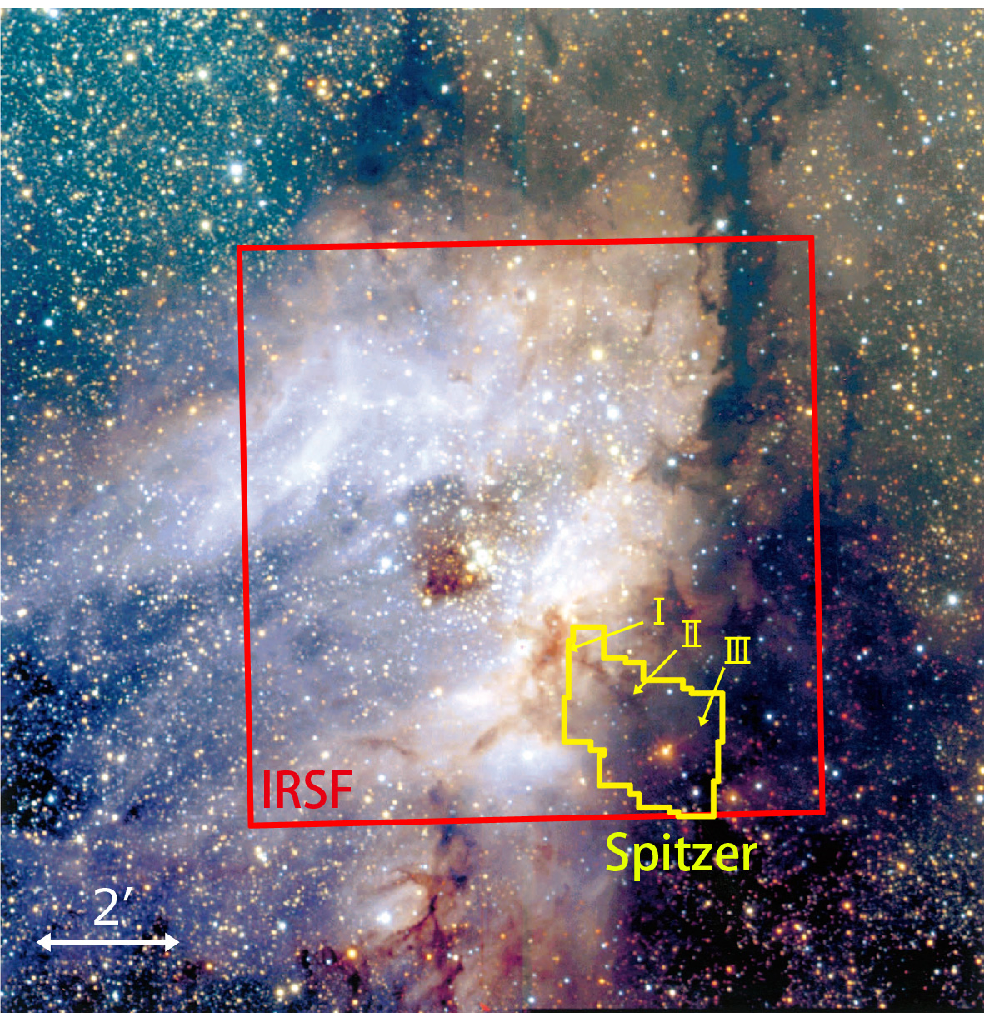}
\caption{Regions observed with $Spitzer$ and IRSF overlaid on the near-IR three color image of M17 (\citealt{Jiang02}). I--III indicate the positions where the mid-IR spectra in Fig.~\ref{pahfit_result_matome} are extracted.}
\label{jiang_area}
\end{figure}

\begin{figure}
\epsscale{0.45}
\plotone{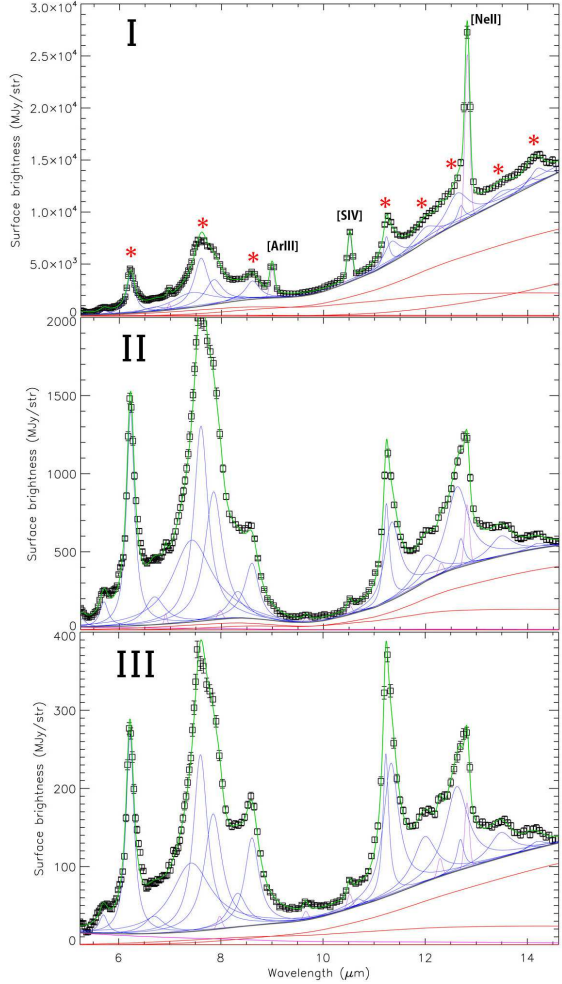}
\caption{Examples of the mid-IR spectra extracted from the areas of $3\arcsec.6\times3\arcsec.6$, which are named as positions I--III in Fig.~{\ref{jiang_area}}. Red asterisks indicate PAH features at wavelengths of 6.2, 7.7, 8.6, 11.3, 12.0, 12.7, 13.5, and 14.2~$\micron$. The green curve indicates the best-fit model using PAHFIT (\citealt{Smith07a}), while the blue, magenta, and red curves indicate the best-fit components of the PAH features, fine-structure lines, and continuum emission, respectively.}
\label{pahfit_result_matome}
\end{figure}

Figure~\ref{pahfit_result_matome} shows examples of the spectra extracted from positions I, II, and III which are labeled in Fig.~\ref{jiang_area}.
The spectra show a variety of emission features due to PAHs at 6.2, 7.7, 8.6, 11.3, 12.0, 12.7, 13.5, and 14.2~$\micron$, and fine-structure lines ([NeII], [ArIII], and [SIV]).
The fine-structure lines are dominant at position I which is nearest to the M17 center, while the PAH features are dominant at positions II and III.
In order to examine variations of the PAH intensities and interband ratios, we decomposed the spectra using PAHFIT (\citealt{Smith07a}), which is designed to fit various spectral features in $Spitzer$/IRS SL spectra including silicate dust absorption around 9.7~$\micron$.
We used PAHFIT assuming screen dust extinction, and reconstructed the spectral maps of the mid-IR features.
In the mid-IR spectra, the PAH~7.7~$\micron$, 11.3~$\micron$, and 12.7~$\micron$ features are treated as PAH complexes which have multiple components.
As treated in \citet{Smith07a}, we calculated a sum of the three components centered at 7.42~$\micron$, 7.60~$\micron$, and 7.85~$\micron$ for the intensity of the PAH~7.7~$\micron$ complex, the two components centered at 11.23~$\micron$ and 11.33~$\micron$ for the intensity of the PAH~11.3~$\micron$ complex, and the two components centered at 12.62~$\micron$ and 12.69~$\micron$ for the intensity of the PAH~12.7~$\micron$ complex.
In the fitting procedure, we considered the systematic error of 2~\% (\citealt{Lebouteiller11}) in addition to random errors provided by CUBISM.
Figure~\ref{pahfit_result_matome} also shows the results of PAHFIT, where PAHFIT reproduces the example spectra very well.

\begin{figure}
\epsscale{1.0}
\plotone{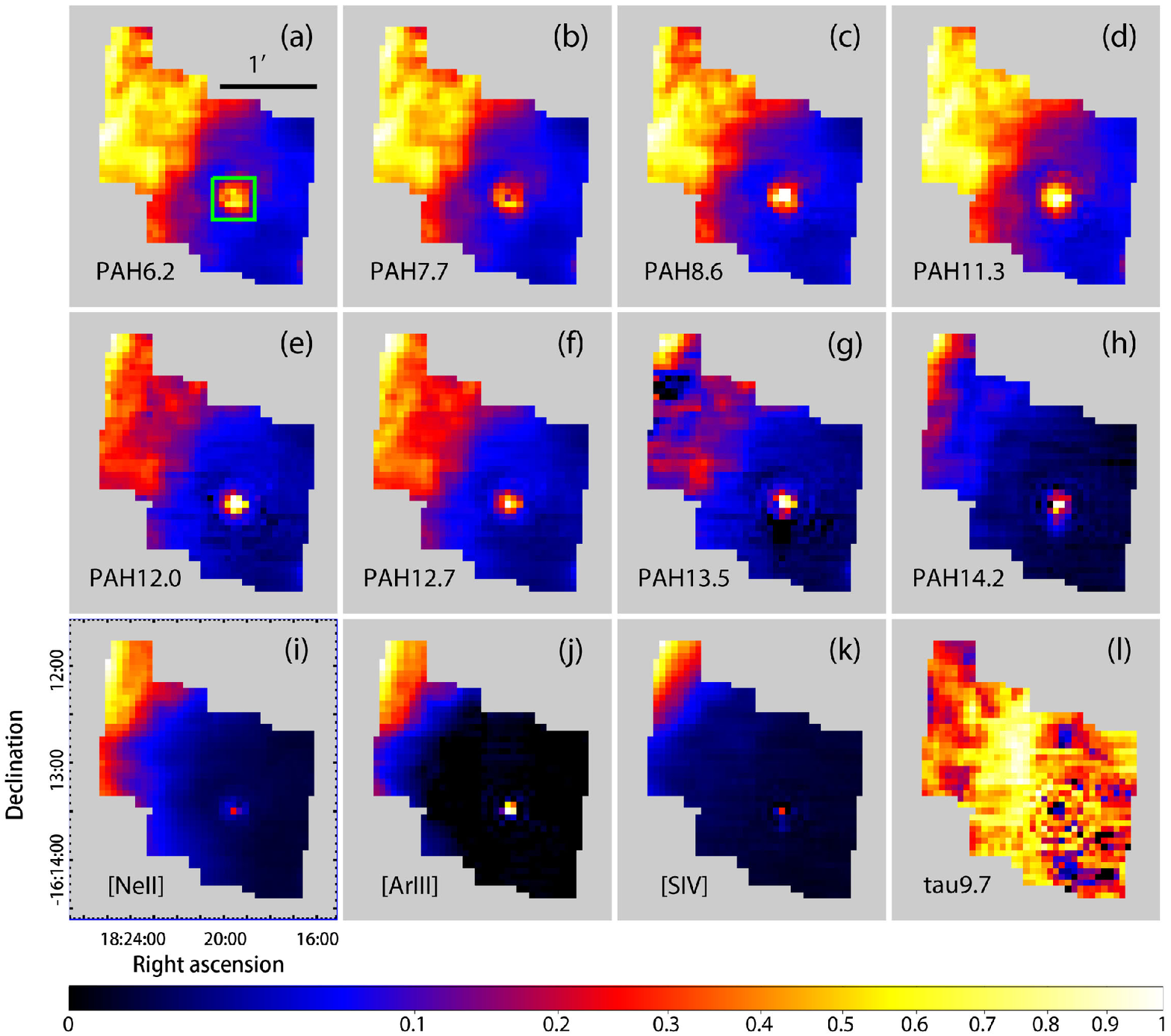}
\caption{Spectral maps of the (a) PAH 6.2~$\micron$, (b) PAH 7.7~$\micron$, (c) PAH 8.6~$\micron$, (d) PAH 11.3~$\micron$, (e) PAH 12.0~$\micron$, (f) PAH 12.7~$\micron$, (g) PAH 13.5~$\micron$, (h) PAH 14.2~$\micron$ features, the (i) [NeII], (j) [ArIII], and (k) [SIV] lines, and (l) optical depth of silicate dust absorption at 9.7~$\micron$ obtained with PAHFIT (\citealt{Smith07a}). The maps are normalized by the following maximum values: (a) $1.2\times10^{-4}~\mathrm{W/m^{2}/sr}$, (b) $3.8\times10^{-4}~\mathrm{W/m^{2}/sr}$, (c) $6.2\times10^{-5}~\mathrm{W/m^{2}/sr}$, (d) $6.0\times10^{-5}~\mathrm{W/m^{2}/sr}$, (e) $3.7\times10^{-5}~\mathrm{W/m^{2}/sr}$, (f) $7.2\times10^{-5}~\mathrm{W/m^{2}/sr}$, (g) $2.6\times10^{-5}~\mathrm{W/m^{2}/sr}$, (h) $2.6\times10^{-5}~\mathrm{W/m^{2}/sr}$, (i) $4.5\times10^{-5}~\mathrm{W/m^{2}/sr}$, (j) $2.3\times10^{-5}~\mathrm{W/m^{2}/sr}$, (k) $3.6\times10^{-5}~\mathrm{W/m^{2}/sr}$, and (l) 1.5. The area indicated by the green box in panel (a) includes a well known young stellar object, where the spectra show ice absorption features and are not fitted well with PAHFIT.
}
\label{specmap}
\end{figure}

Figure~\ref{specmap} shows the resultant spectral maps obtained for the mid-IR features (PAH features at 6.2, 7.7, 8.6, 11.3, 12.0, 12.7, 13.5, and 14.2~$\micron$, [NeII], [ArIII], [SIV], and optical depth of silicate dust absorption at 9.7~$\micron$), where the interstellar extinction is corrected.
In the maps, there is a large-scale gradient from the north-east to the south-west.
In the PAH maps, there are also local structures especially on the east side, suggesting that PAHs are not only globally but also locally affected by the surrounding interstellar environments.
It is notable that the maps of the PAH 12.7~$\micron$ and 12.0~$\micron$ features are similar to each other.
They are both probably attributed to C-H out-of-plane bending modes, and the former is blended with the strong [NeII] line, while the latter is not.
Therefore, this similarity suggests that PAHFIT has successfully decomposed emissions from the PAH 12.7~$\micron$ feature and the [NeII] line.
In the fine-structure line maps, a spatial extent in the [NeII] map is large, while that in [SIV] is small, which presumably corresponds to the difference in the ionization potential; the energies required to ionize Ne$^0$, Ar$^+$, and S$^{++}$ are 21.6~eV, 27.6~eV, and 34.8~eV, respectively.
In the optical depth map, there is a ridge structure in the north-south direction.
We confirm that the optical depth estimated by using PAHFIT is consistent with that estimated in \citet{Stock16}.

\begin{figure}
\epsscale{.45}
\plotone{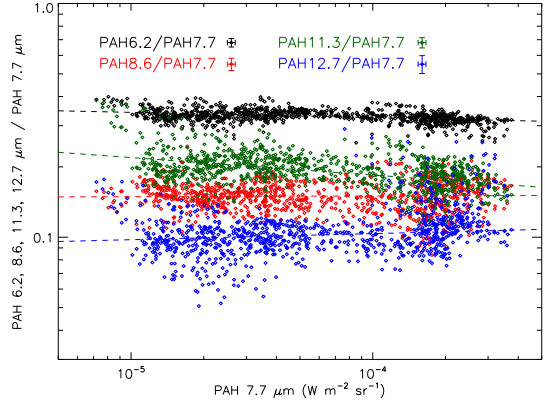}
\caption{Interband ratios of the PAH 6.2~$\micron$, 8.6~$\micron$, 11.3~$\micron$, and 12.7~$\micron$ features to the PAH 7.7~$\micron$ feature plotted against intensities of the PAH 7.7~$\micron$ feature. The median value of the uncertainties for each interband ratio is shown as a typical error bar in the figure legend. The dashed lines indicate the best-fit power-law relation for each interband ratio. }
\label{pah_correlation}
\end{figure}

Figure~\ref{pah_correlation} shows the PAH interband ratios of the PAH 6.2~$\micron$, 8.6~$\micron$, 11.3~$\micron$, and 12.7~$\micron$ features to the PAH 7.7~$\micron$ feature plotted against intensities of the PAH 7.7~$\micron$ feature.
The dashed lines in Fig.~\ref{pah_correlation} indicate the best-fit power-law relation for each of the PAH interband ratio.
It is notable that the best-fit lines are nearly flat, indicating that there are no clear systematic variations in the PAH interband ratios for various interstellar environments.
By contrast, there are clear differences in the dispersion along the vertical axis from ratio to ratio.
We calculated 1$\sigma$ standard deviations of the PAH interband ratios.
As a result, we find that the PAH 6.2~$\micron$/PAH 7.7~$\micron$ ratios show the smallest variations (PAH 6.2~$\micron$/PAH 7.7~$\micron$: 0.028~dex (decimal exponent), PAH 8.6~$\micron$/PAH 7.7~$\micron$: 0.065~dex, PAH 11.3~$\micron$/PAH 7.7~$\micron$: 0.086~dex, PAH 12.7~$\micron$/PAH 7.7~$\micron$: 0.11~dex).
Probable origins of the PAH 6.2~$\micron$ and 7.7~$\micron$ features are C-C bonds, while those of the other PAH features are C-H bonds (\citealt{Allamandola89, Draine07}).
Therefore, the difference in the variations in the interband ratios may reflect spatial variations in the properties of the PAH population.

\begin{figure}
\epsscale{.45}
\plotone{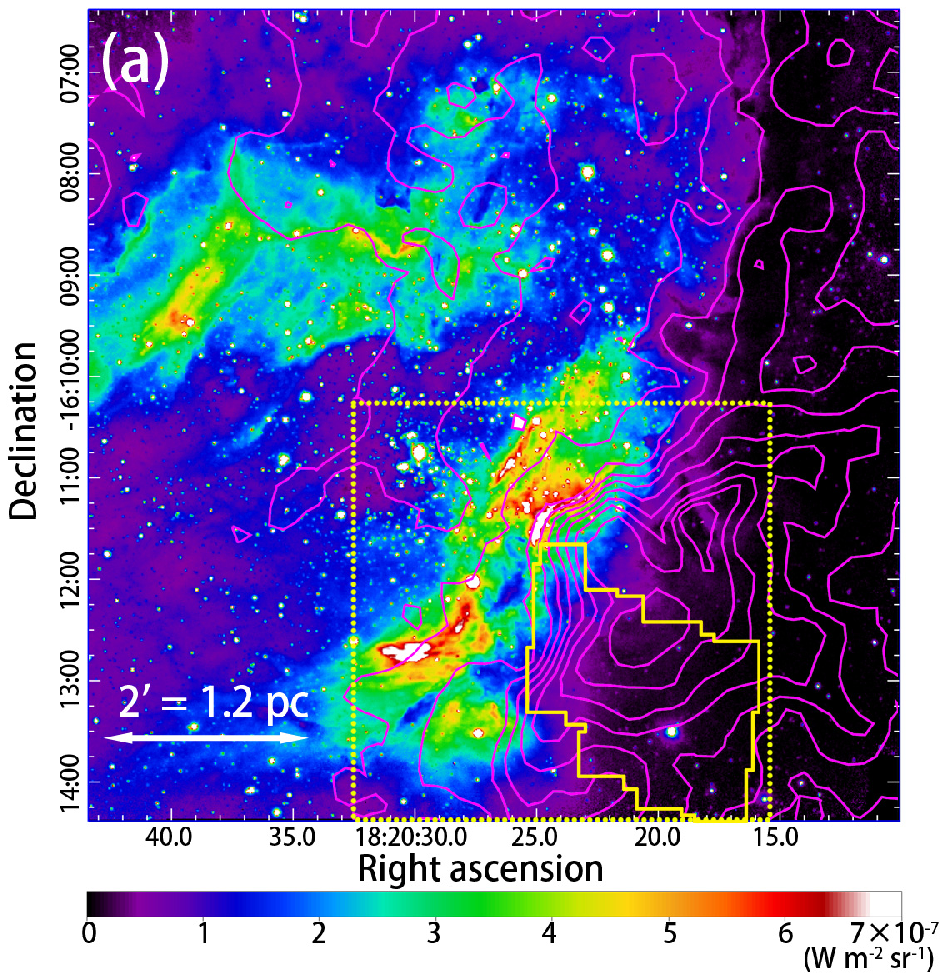}
\epsscale{.4}
\plotone{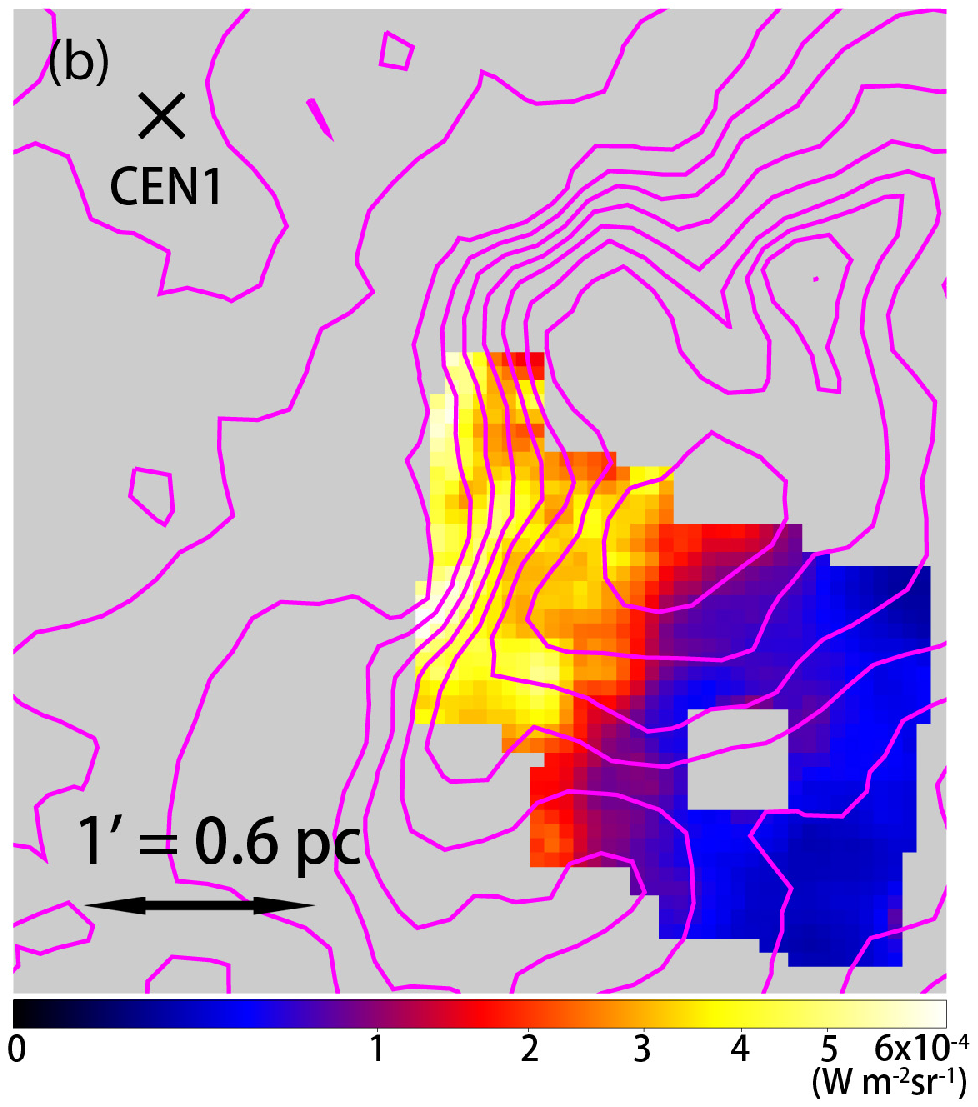}
\caption{(a) Br$\gamma$ narrow-band image of M17 taken with IRSF, where the contours are the $^{13}$CO ($J$=1-0) intensity integrated over the velocity range of 10--30 km/s. The contours are drawn at ten linearly spaced levels from the 5$\sigma$ detection level of 20 K~km/s to the peak of 206 K~km/s. The yellow box indicates the observed region with $Spitzer$/IRS. The area indicated by the yellow dotted box is enlarged in panel (b). (b) Total intensity map of the major PAH features (PAH~6.2~$\micron$, PAH~7.7~$\micron$, PAH~8.6~$\micron$, PAH~11.3~$\micron$, and PAH~12.7~$\micron$) overlaid on the $^{13}$CO ($J$=1-0) contour map. The cross represents the position of the most prominent O4-type star in M17, CEN1. The region indicated by the green box in Fig.~\ref{specmap}(a) is masked.}
\label{Brg_CO_map}
\end{figure}

Figure~\ref{Brg_CO_map}(a) shows the $^{13}$CO($J$=1-0) contour map overlaid on the Br$\gamma$ image of M17.
The Br$\gamma$ emission is detected in an inner region of M17, especially in south-west and north-east areas, but not detected in west and south-west areas.
In the regions where Br$\gamma$ is not detected, the $^{13}$CO emission is strongly detected.
The strong peak of the $^{13}$CO emission is adjacent to the bright rim of the Br$\gamma$ emission.
Thanks to the high spatial resolution of the $^{13}$CO map and the small extinction in the Br$\gamma$ map, a spatial separation of an HII region from the molecular cloud is clearly recognized.
Figure~\ref{Brg_CO_map}(b) shows comparison of the total major PAH map (PAH 6.2~$\micron$+PAH 7.7~$\micron$+PAH 8.6~$\micron$+PAH 11.3~$\micron$+PAH 12.7~$\micron$) with the $^{13}$CO map.
By comparing Fig.~\ref{Brg_CO_map}(a) with Fig.~\ref{Brg_CO_map}(b), it is clear that the PAH emission is bright on the boundary between the Br$\gamma$ and $^{13}$CO emissions, which strongly supports that the PAH features are emitted from PDRs.

\begin{figure}
\epsscale{1.0}
\plotone{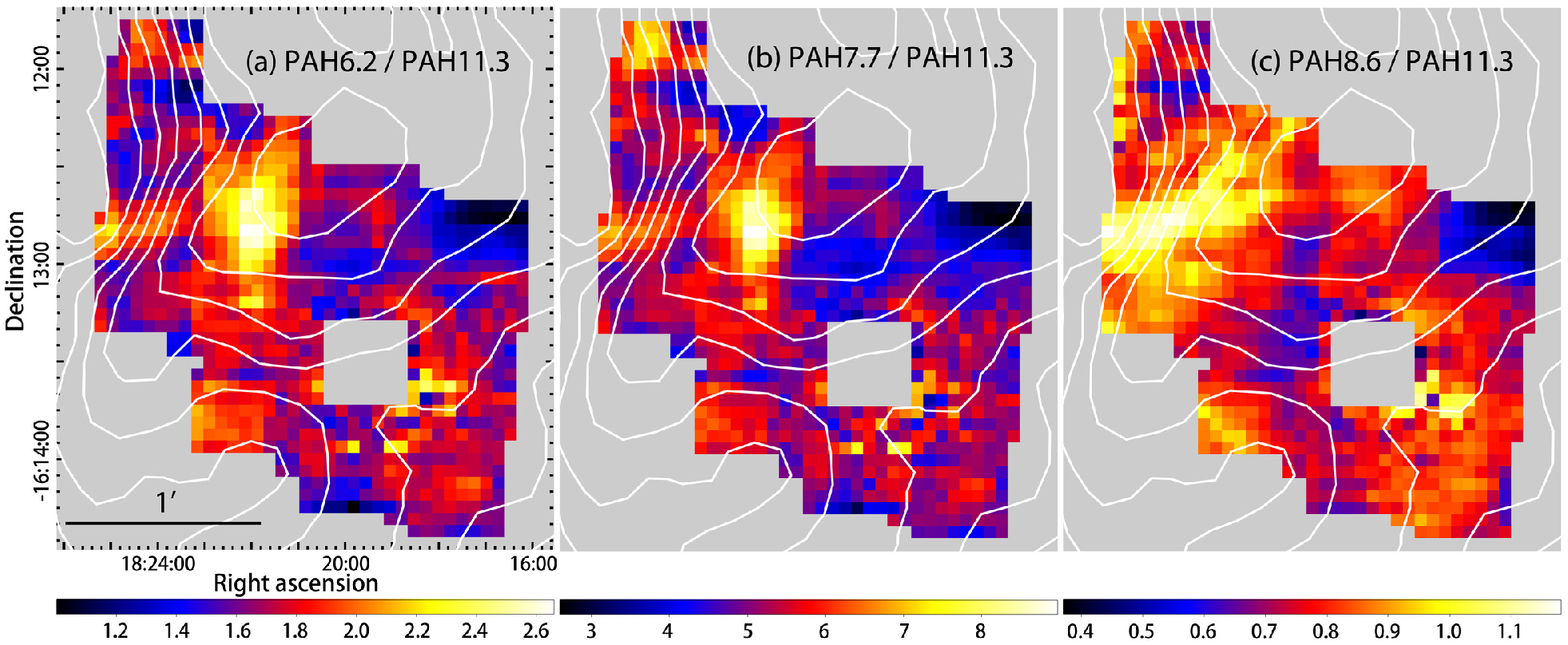}\\
\epsscale{0.48}
\plotone{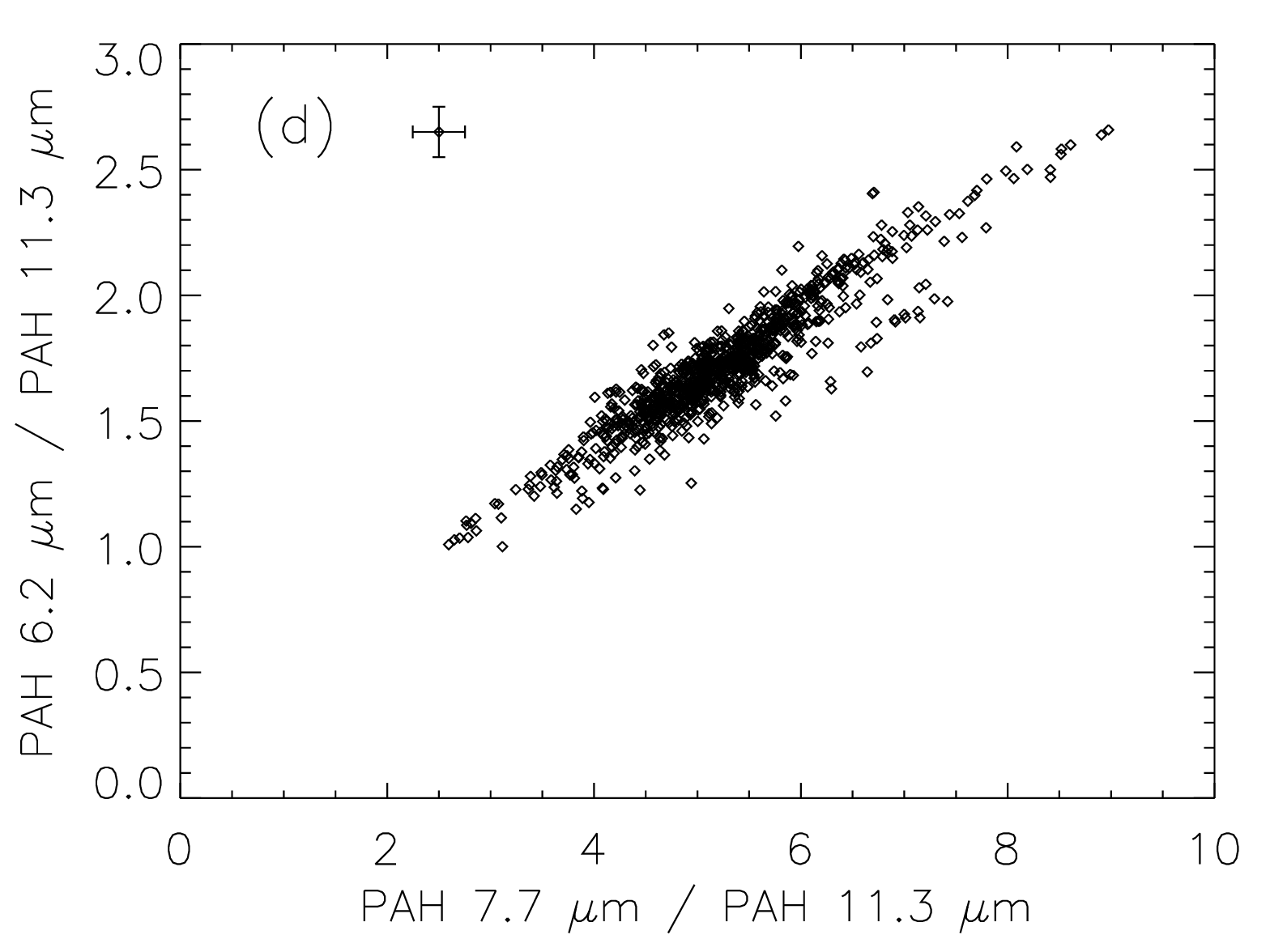}
\plotone{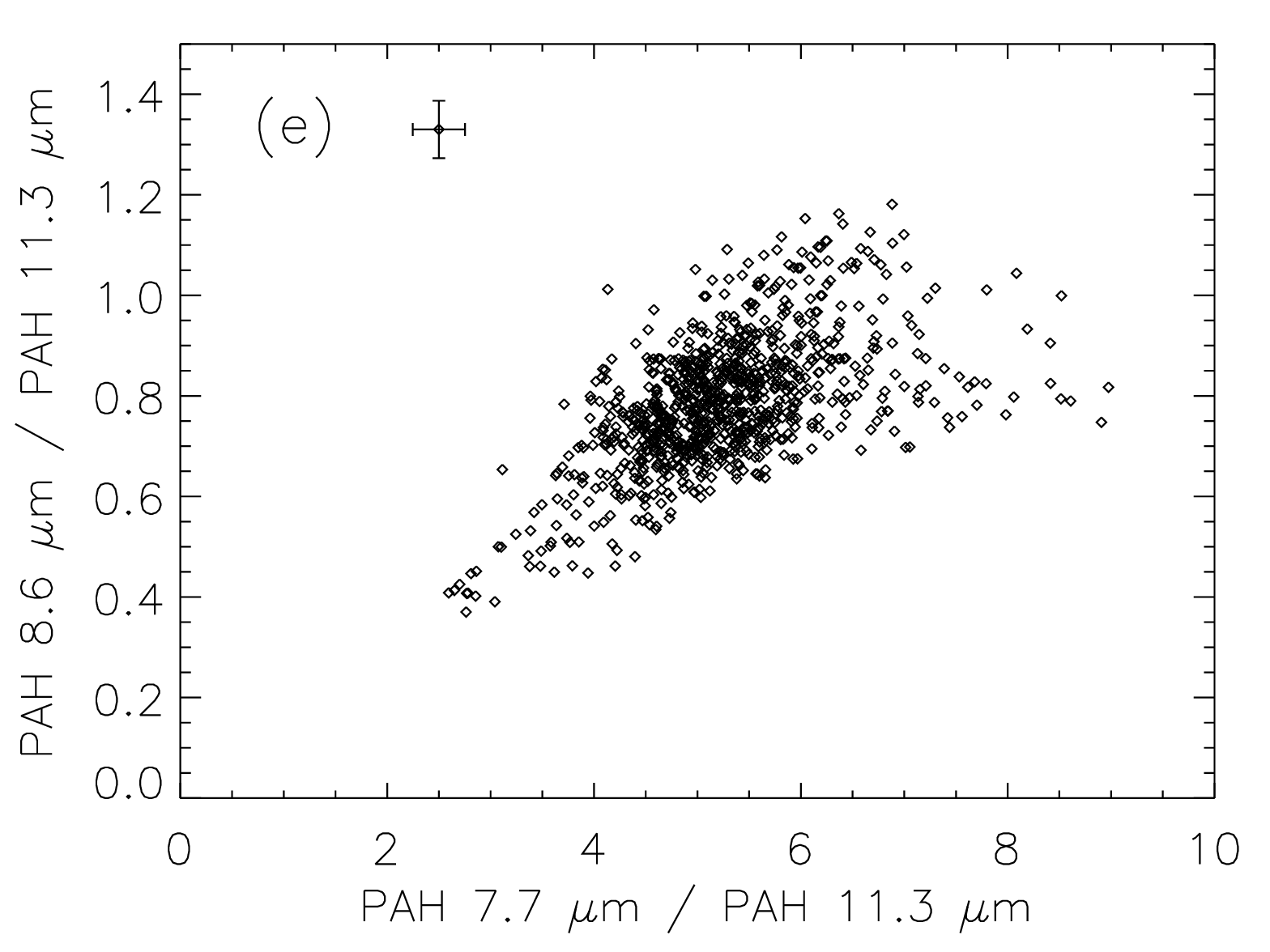}
\caption{(a-c) Interband ratio maps of the PAH 6.2~$\micron$, 7.7~$\micron$, and 8.6~$\micron$ features to the PAH 11.3~$\micron$ feature. The contours are the same as shown in Fig.~\ref{Brg_CO_map}. The region indicated by the green box in Fig.~\ref{specmap}(a) is masked. (d, e) Correlation plots between the above interband ratios. A typical error is shown on the upper left corner.}
\label{othermap}
\end{figure}

Figure~\ref{othermap} shows the PAH interband ratio maps (PAH 6.2~$\micron$/PAH 11.3~$\micron$, PAH 7.7~$\micron$/PAH 11.3~$\micron$, and PAH 8.6~$\micron$/PAH 11.3~$\micron$) and their correlation plots.
The interband ratios are possible probes of the degree of PAH ionization because the PAH 6.2, 7.7, and 8.6~$\micron$ features are efficiently emitted by ionized PAHs, while the PAH 11.3~$\micron$ feature is emitted by both ionized and neutral PAHs (\citealt{Allamandola99, Peeters02}).
In the maps, there is no clear large-scale gradient as seen in Fig.~\ref{specmap}, and local structures are dominant, suggesting that the PAH properties are controlled by local conditions rather than the large-scale UV environment determined by the OB stars in M17.
It is notable that these ratios are high near the $^{13}$CO peak.
Position II in Fig.~\ref{jiang_area} corresponds to the peak of the PAH 7.7~$\micron$/PAH 11.3~$\micron$ ratio.
The spectrum extracted from position II actually shows a high PAH 7.7~$\micron$/PAH 11.3~$\micron$ ratio in comparison with those extracted from positions I and III (Fig.~\ref{pahfit_result_matome}).
This result supports that the degree of PAH ionization is significantly different from position to position.
The PAH 6.2~$\micron$/PAH 11.3~$\micron$ ratios tightly correlate with the PAH 7.7~$\micron$/PAH 11.3~$\micron$ ratios, while they do not tightly correlate with the PAH 8.6~$\micron$/PAH 11.3~$\micron$ ratios, although the PAH 8.6~$\micron$/PAH 11.3~$\micron$ ratio is also known to be a measure of the PAH ionization (e.g., \citealt{Boersma14}).
Since the probable origin of the PAH 8.6~$\micron$ feature (C-H in-plane-bending mode; \citealt{Tielens05}) is different from that of the PAH 6.2~$\micron$ and 7.7~$\micron$ features (C-C stretching mode; \citealt{Tielens05}), other factors (e.g., size) may affect the PAH 8.6~$\micron$/PAH 11.3~$\micron$ ratio (e.g., \citealt{Bauschlicher08, Bauschlicher09, Ricca12}).
Maps and correlation plots of other four interband ratios (PAH 12.0~$\micron$/PAH 11.3~$\micron$, PAH 12.7~$\micron$/PAH 11.3~$\micron$, PAH 13.5~$\micron$/PAH 11.3~$\micron$, and PAH 14.2~$\micron$/PAH 11.3~$\micron$) are shown in Fig.~\ref{othermap2}, which are likely probes of the PAH edge structure.
The maps show that these ratios are relatively high on the near side of M17, suggesting that PAH structures are not uniform in the region.
The plots show global correlations among these interband ratios, supporting the same origin for these features (i.e., C-H out-of-plane bending mode).

\begin{figure}
\epsscale{0.75}
\plotone{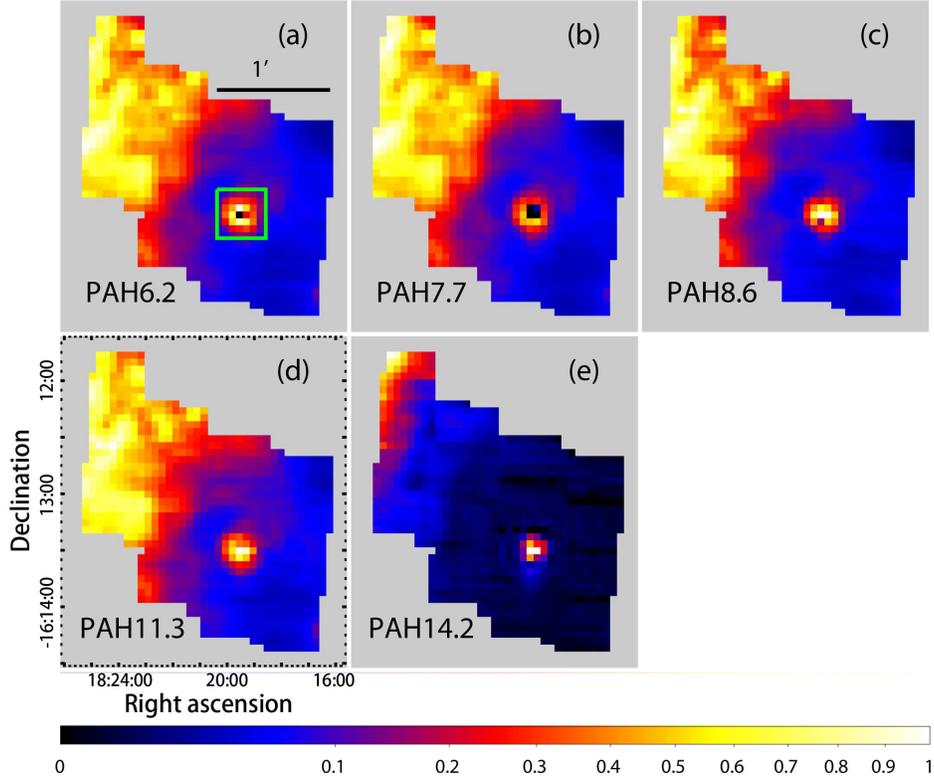}
\caption{Spectral maps of the (a) PAH 6.2~$\micron$, (b) PAH 7.7~$\micron$, (c) PAH 8.6~$\micron$, (d) PAH 11.3~$\micron$, and (e) PAH 14.2~$\micron$ features obtained with the linear-baseline-fitting method. The maps are normalized by the following maximum values: (a) $6.5\times10^{-5}~\mathrm{W/m^{2}/sr}$, (b) $1.3\times10^{-4}~\mathrm{W/m^{2}/sr}$, (c) $1.4\times10^{-5}~\mathrm{W/m^{2}/sr}$, (d) $3.0\times10^{-5}~\mathrm{W/m^{2}/sr}$, and (e) $1.5\times10^{-5}~\mathrm{W/m^{2}/sr}$. The area indicated by the green box in Fig.~\ref{specmap}(a) is also shown in panel (a).
}
\label{robust_intensity}
\end{figure}

\begin{figure}
\epsscale{1.0}
\plotone{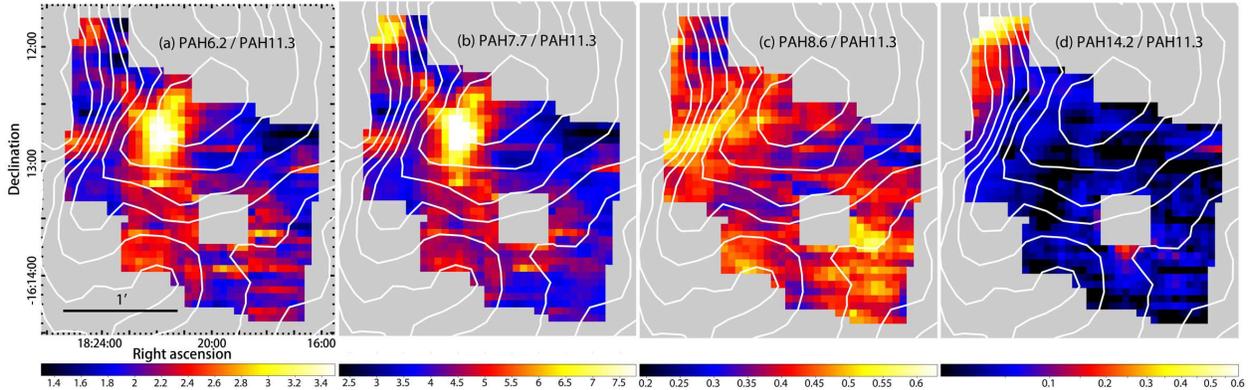}
\caption{Examples of the PAH interband ratio maps derived with the linear-baseline-fitting method. The region indicated by the green box in Fig.~\ref{specmap}(a) is masked. The contours are the same as shown in Fig.~\ref{Brg_CO_map}.
}
\label{robust_ratio}
\end{figure}

In order to check robustness of the above results, we also evaluated the PAH features in the following model-independent manner: we used linear baselines of 5.8--6.5~$\micron$, 7.1--8.3~$\micron$, 8.3--8.8~$\micron$, 10.8--11.8~$\micron$, and 13.8--14.5~$\micron$ for the PAH 6.2~$\micron$, 7.7~$\micron$, 8.6~$\micron$, 11.3~$\micron$, and 14.2~$\micron$ features, respectively, to estimate their intensities.
Figures~\ref{robust_intensity}, \ref{robust_ratio}(a)-\ref{robust_ratio}(c), and \ref{robust_ratio}(d) correspond to Figs.~\ref{specmap}, \ref{othermap}(a)-\ref{othermap}(c), and \ref{othermap2}(d), as obtained in this method.
We confirm from these figures that the linear-baseline-fitting results show overall consistency with the PAHFIT result.
\citet{GallianoF08} showed that the PAH interband ratios are mostly independent of the approach isolating the PAH features.

\section{Discussion}

\subsection{Spatial variations in the degree of PAH ionization}

We find that the PAH 7.7~$\micron$/PAH 11.3~$\micron$ ratio ranges from 2.6 to 9.0 with a median value of 5.2, suggesting that the degree of PAH ionization is significantly variable in M17SW.
In order to derive a map of the degree of PAH ionization, we estimate the degree of ionization from the PAH 7.7~$\micron$/PAH 11.3~$\micron$ ratio using the following equation in \citet{Joblin96}:
\begin{equation}
\frac{I_{7.7}}{I_{11.3}} = \frac{{I^0}_{7.7}}{{I^0}_{11.3}} 
                           \frac{\left( 1+ \frac{n^+}{n^0}\frac{{I^+}_{7.7}}{{I^0}_{7.7}} \right)}
                                {\left( 1+ \frac{n^+}{n^0}\frac{{I^+}_{11.3}}{{I^0}_{11.3}} \right)},
\end{equation}
where $I_{7.7}$, $I_{11.3}$, and $n$ indicate the intensities of the PAH 7.7~$\micron$ and PAH 11.3~$\micron$ features, and the number density of PAHs, respectively.
The zero and plus in the superscript indicate the neutral and ionized states, respectively.
We assumed ${I^0}_{7.7}/{I^0}_{11.3}$, ${I^+}_{7.7}/{I^0}_{7.7}$, and ${I^+}_{11.3}/{I^0}_{11.3}$ to be 1.3, 5.54, and 0.6, respectively (\citealt{Fleming10}), and converted the observed $I_{7.7}/I_{11.3}$ to $n^+/n^0$.
Finally, we obtained the degree of PAH ionization of $(1 +n^0/n^+)^{-1}$.

\begin{figure}
\epsscale{0.4}
\plotone{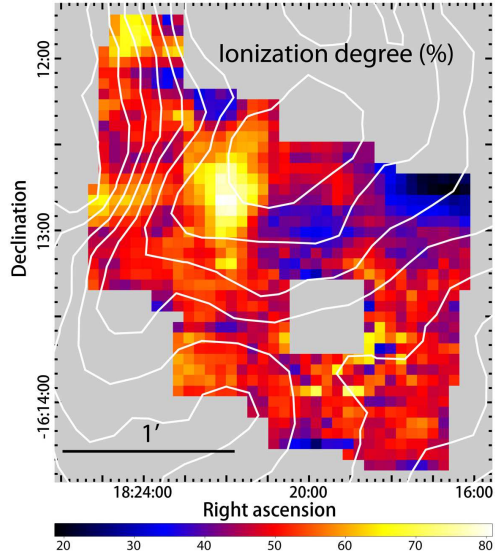}
\caption{Degree of PAH ionization estimated from the PAH 7.7~$\micron$/PAH 11.3~$\micron$ ratios. The contours are the same as shown in Fig.~\ref{Brg_CO_map}. The region indicated by the green box in Fig.~\ref{specmap}(a) is masked.}
\label{ionization_degree}
\end{figure}

Figure~\ref{ionization_degree} shows a map of the degree of PAH ionization thus derived.
In the figure, the degree of PAH ionization ranges from 19~\% to 81~\% with a median value of 48~\%.
These values correspond to the ionization parameter, $G_0\sqrt{T_{\mathrm{gas}}}/n_e$, of 9$\times10^3$ to 1$\times10^5$ (\citealt{Tielens05}), where $G_0$, $T_{\mathrm{gas}}$, and $n_e$ indicate the far-UV intensity normalized by that of the solar neighborhood, gas temperature, and electron density, respectively, and we assumed the number of carbon atoms in a PAH to be 50.
Therefore, the degree of PAH ionization varies in a wide range from region to region, and typical PAHs are moderately ionized in M17SW.
Such a situation is similar to the cases of the previous studies; the degree of PAH ionization is 10--70~\% and typically 50~\% in NGC~7023 (\citealt{Boersma13}), while that is 20--100~\% and typically 50~\% in the Orion Bar (\citealt{Haraguchi12}), although the approaches to determine the degree of PAH ionization are different between those studies and the present one.
The minimum degree of PAH ionization of $\sim$15~\% is common among the three objects, suggesting that PAHs in molecular clouds adjacent to star-forming regions are slightly ionized.
The maximum degree of PAH ionization of 80\% in M17SW is lower than that of Orion Bar and comparable to that in NGC~7023.
This difference is presumably due to the difference in the observed positions; the exciting stars are covered in the observational area of Orion Bar, while those are not covered in those of M17 and NGC~7023.

\begin{figure}
\epsscale{0.4}
\plotone{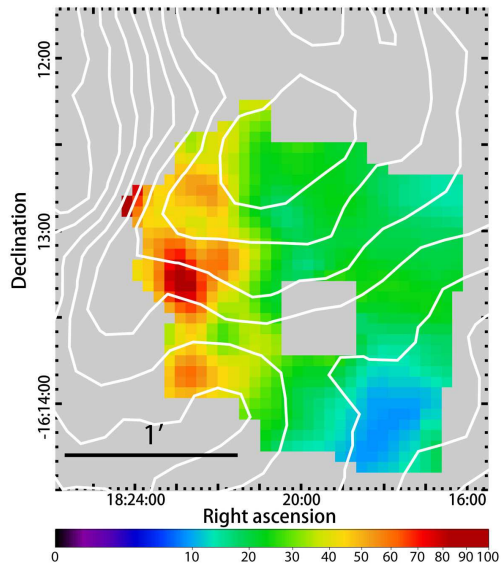}
\caption{Distribution of $G_0$ estimated from the Herschel/PACS 70 and 100~$\micron$ maps. The color bar is shown in the square root scale. The spatial resolution of the map is 6.$\arcsec$7 which is the PSF size of the PACS 100~$\micron$ map. The contours are the same as shown in Fig.~\ref{Brg_CO_map}. The regions showing strong [NeII] ($>5.0\times10^{-6}~\mathrm{W/m^{2}/sr}$) are masked.}
\label{ionization_parameter}
\end{figure}

One interesting result in Fig.~\ref{ionization_degree} is that the degree of PAH ionization is high near the peak of the $^{13}$CO emission.
As expressed in the definition of the ionization parameter, the balance between the ionization and recombination is important in the degree of PAH ionization.
The ionization depends on the far-UV radiation field, $G_0$, while the recombination mainly depends on the electron density, $n_e$.
Therefore, we evaluate $G_0$ and $n_e$ to verify the spatial variation of the degree of PAH ionization. 
We determined $G_0$ with the Herschel/PACS 70 and 100~$\micron$ maps which we retrieved from NASA/IPAC Infrared Science Archives (Observation ID: 1342192767).
We converted a 70~$\micron$/100~$\micron$ color temperature to $G_0$ assuming the power-law dust emissivity index of $\beta$=1 and the relation between $G_0$ and $T_\mathrm{dust}$, $G_0 = (T_\mathrm{dust}/17.5)^5$ (\citealt{Boulanger96}). 
The derived $G_0$ map is shown in Fig.~\ref{ionization_parameter}, where obvious HII regions show strong [NeII] ($>5.0\times10^{-6}~\mathrm{W/m^2/sr}$) are masked out as the conversion is not applicable to HII regions.
Electrons in PDRs are mainly produced by ionization of neutral carbon atoms.
Therefore, [CII] is a possible probe of $n_e$ for a nearly constant gas temperature; a [CII] map of this region is obtained by \citet{Perez-Beaupuits12}.

In Fig.~\ref{ionization_parameter}, we find that the $G_0$ map exhibits a local maximum near the peak of the degree of PAH ionization, whereas the [CII] map in \citet{Perez-Beaupuits12} does not have a local minimum there.
These results support the high degree of PAH ionization inside the molecular cloud, which is driven by high $G_0$ rather than low $n_e$.
Although M17SW is one of the most intensively studied regions in our Galaxy, the spatial structure in the line of sight is still unclear; \citet{Stutzki90} and \citet{Sheffer13} proposed clumpy and bowl-shaped structures in M17SW, respectively.
Therefore the local peak of the degree of PAH ionization may be due to geometrical effects, and in reality, this region can be located closely to the M17 center.
This situation is, however, rather unlikely, because Fig.~\ref{othermap2} shows no corresponding local peak in the interband ratios; such a region should also show variations in the PAH structure due to the intense UV radiation from the M17 center.
It should be noted that the region with high PAH ionization degree is associated with dense gas probed by $^{13}$CO.
Additionally, since PAHs are ionized by far-UV photons, B-type or later stars can substantially ionize PAHs.
In this situation, such stars may be reasonable ionizing source candidates because the Br$\gamma$ emission is not locally enhanced in the region (Fig.~\ref{Brg_CO_map}).
Therefore, it is possible that B-type or later stars buried inside the molecular cloud locally ionize PAHs.
As can be seen in Fig.~\ref{ionization_parameter}, an even stronger $G_0$ peak is found at $\sim30\arcsec$ to the south, where the degree of PAH ionization does not show a significant increase (Fig.~\ref{ionization_degree}).
Young stellar objects which cannot substantially ionize PAHs may attribute to the $G_0$ peak here.

\subsection{Spatial variations of the PAH structure}

\begin{figure}
\epsscale{1.0}
\plotone{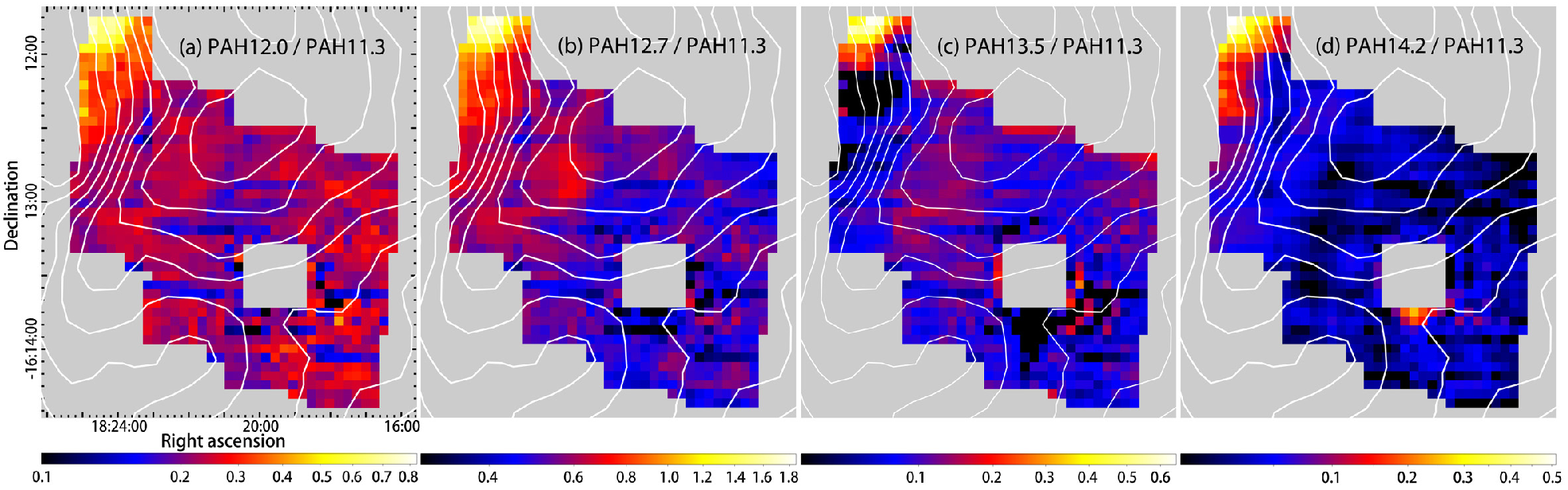}\\
\epsscale{0.23}
\plotone{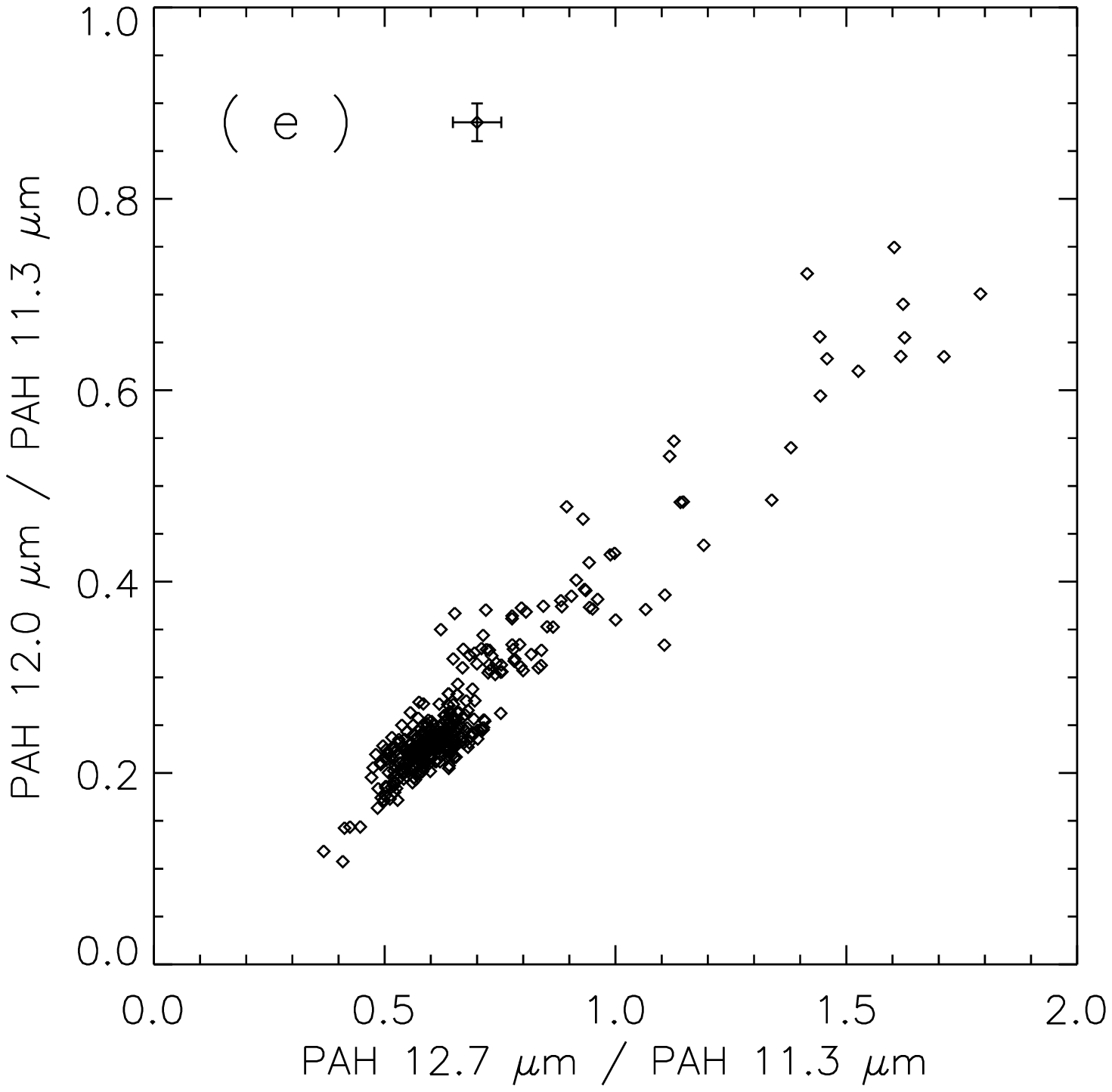}
\plotone{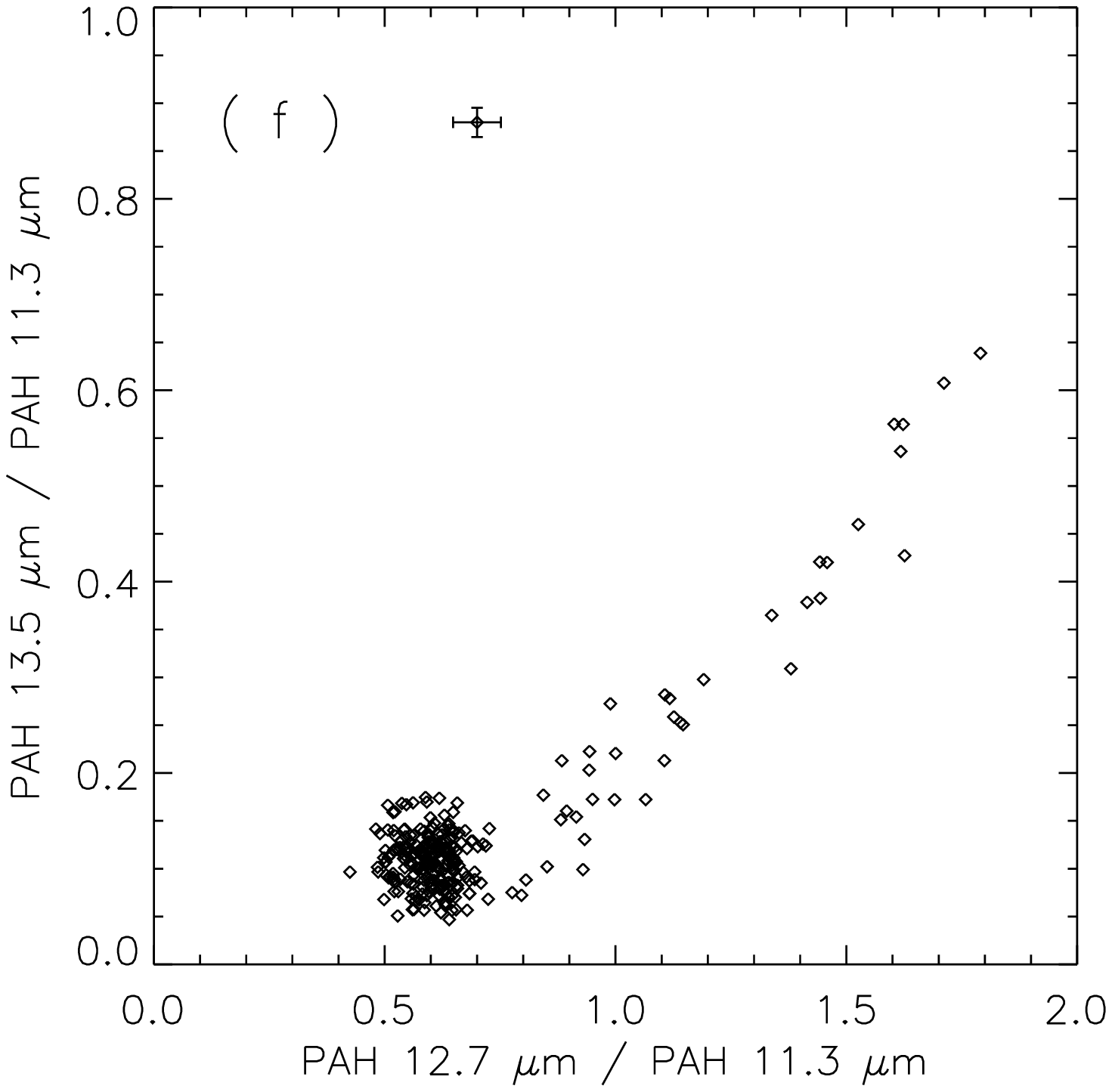}
\plotone{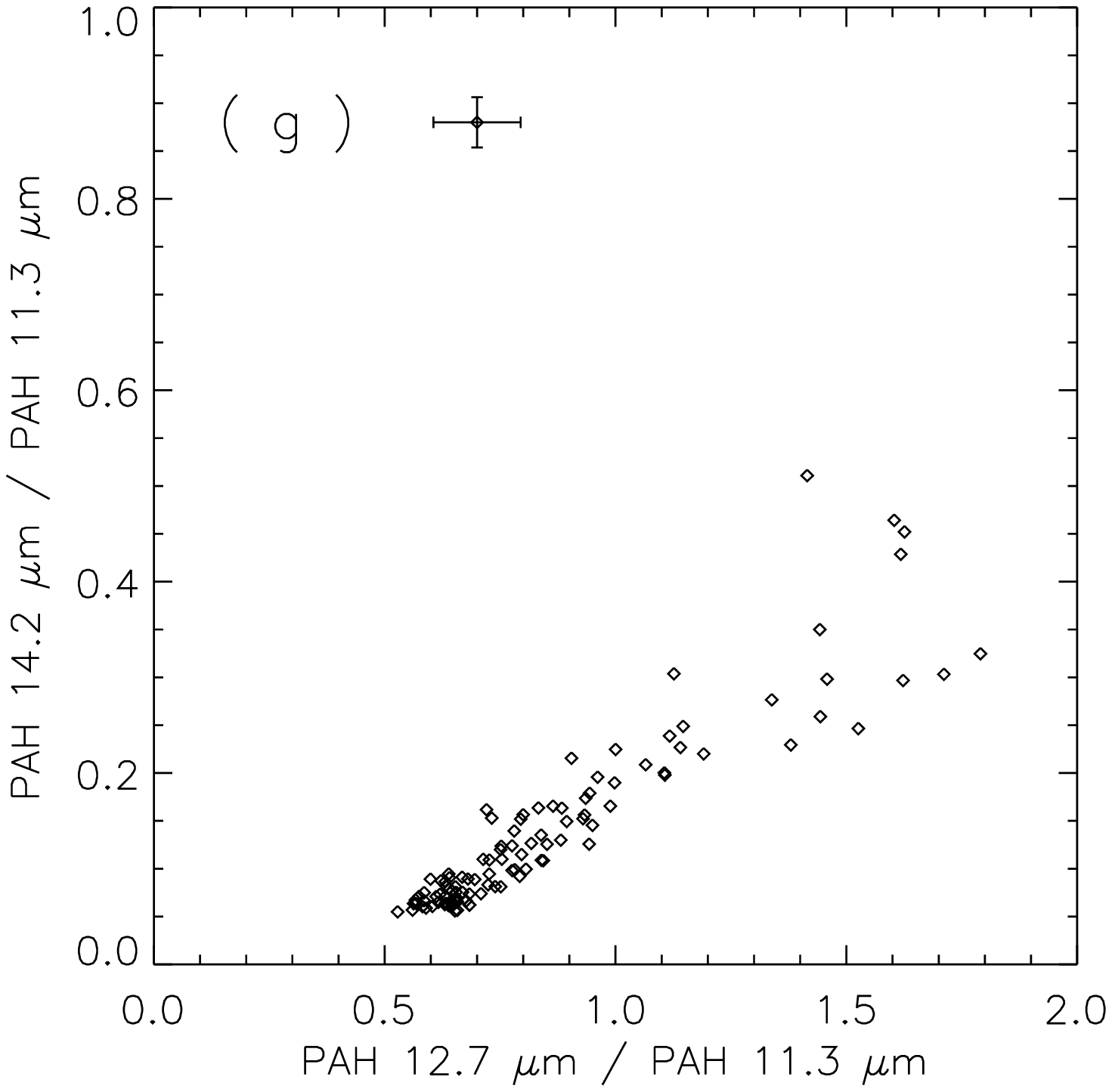}
\plotone{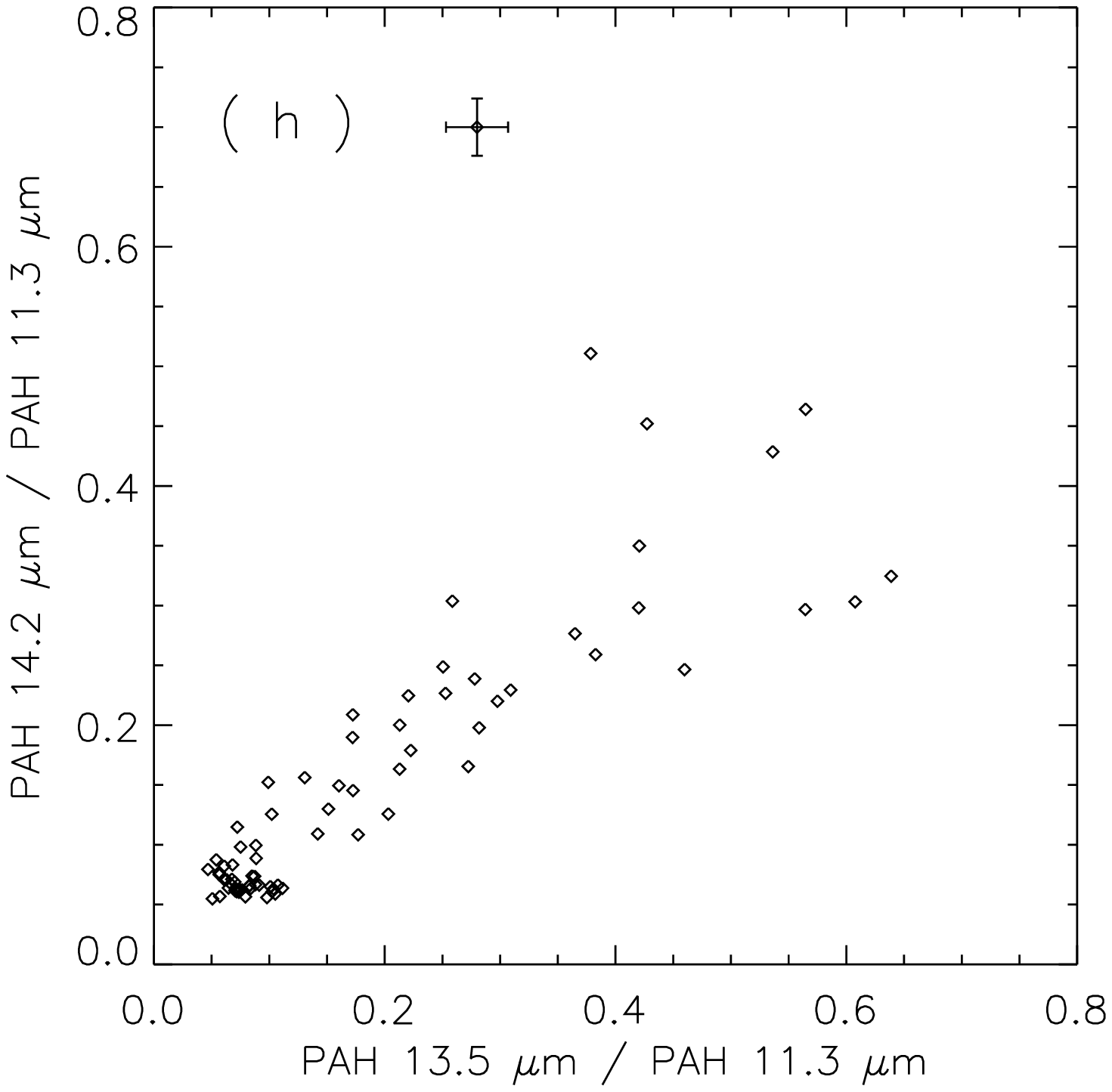}
\caption{(a-d) Interband ratio maps of the PAH 12.0~$\micron$, 12.7~$\micron$, 13.5~$\micron$, and 14.2~$\micron$ features to the PAH 11.3~$\micron$ feature. The contours are the same as shown in Fig.~\ref{Brg_CO_map}. The region indicated by the green box in Fig.~\ref{specmap}(a) is masked. (e-h) Examples of correlation plots between the above interband ratios. The interband ratios are plotted for regions showing strong PAH emissions (more than 10~\% of the peak intensity in Fig.~\ref{specmap}). A typical error is shown on the upper left.}
\label{othermap2}
\end{figure}

In Figs.~\ref{othermap2}(a)-\ref{othermap2}(d), the ratios of the PAH 12.0, 12.7, 13.5, and 14.2~$\micron$ features to the PAH 11.3~$\micron$ feature have a strong peak near the M17 center.
These interband ratios are possible probes of the PAH edge structures.
Likely origins of these features are all C-H out-of-plane bending, but the numbers of adjacent C-H bonds in a benzene ring are different (PAH~11.3~$\micron$: solo; PAH~12.0~$\micron$: duo; PAH~12.7~$\micron$: trio; PAH~13.5~$\micron$ and PAH~14.2~$\micron$: quartet; \citealt{Hony01, Draine03, Tielens05}).
Therefore, Fig.~\ref{othermap2} suggests that the PAH structure is not uniform in the observed region, but is different especially near the M17 center.

Among the interband ratios, the PAH 13.5~$\micron$/PAH 11.3~$\micron$ and the PAH 14.2~$\micron$/PAH 11.3~$\micron$ ratios are more localized toward the center of M17 relative to the PAH 12.0~$\micron$/PAH 11.3~$\micron$ and the PAH 12.7~$\micron$/PAH 11.3~$\micron$ ratios.
Pencil-beam observations by \citet{Hony01} observationally showed that the PAH 13.5~$\micron$/PAH 11.3~$\micron$ and the PAH 14.2~$\micron$/PAH 11.3~$\micron$ ratios in their HII-region samples tend to be higher than those in PDR samples.
Our spectral maps clearly show such spectral variations in M17.
One possible interpretation of the high interband ratios is processing of edgy PAHs.
Since the survival time of edgy PAHs is short, edgy PAHs observed near the M17 center may be freshly released ones from grain mantles due to the intense UV radiation.
Hence Figs.~\ref{othermap2}(a)-\ref{othermap2}(d) may suggest the active PAH erosion/destruction process at the PDR surface.

\begin{figure}
\epsscale{0.4}
\plotone{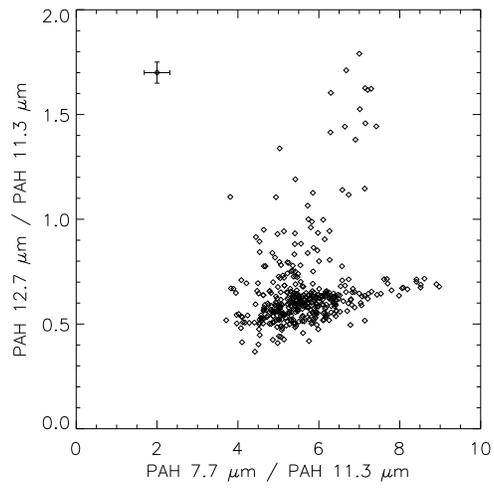}
\caption{PAH 12.7~$\micron$/PAH 11.3~$\micron$ ratios versus the PAH 7.7~$\micron$/PAH 11.3~$\micron$ ratios plotted for regions showing strong PAH emissions (more than 10~\% of the peak intensity in Fig.~\ref{specmap}). A typical error is shown on the upper left.}
\label{ionization_dependence}
\end{figure}

The plots in Figs.~\ref{othermap2}(e)-\ref{othermap2}(h) show overall good correlations between the interband ratios.
\citet{Boersma15} showed that the PAH 12.7~$\micron$/PAH 11.3~$\micron$ ratio depends on both the degree of PAH ionization and the PAH edge structure.
Since the spatial positions showing structural changes and high degree of ionization are different in the present study, our data may be useful to examine the dependence of the edge structure and the degree of PAH ionization on the PAH 12.7~$\micron$/PAH 11.3~$\micron$ ratio.
In Figs.~\ref{othermap2}(e)-\ref{othermap2}(g), the PAH 12.7~$\micron$/PAH 11.3~$\micron$ ratios show good correlation with the ratios probing the PAH edge structure; the correlation coefficients are $R=$0.78 ($N=$399), 0.76 ($N=$284), 0.92 ($N=$111), respectively.
In Fig.~\ref{ionization_dependence}, we also examine the correlation between the PAH 12.7~$\micron$/PAH 11.3~$\micron$ ratios with the PAH 7.7~$\micron$/PAH 11.3~$\micron$ ratios probing the degree of PAH ionization.
In the figure, correlation is relatively weak ($R=$0.28, $N=$418), suggesting that the PAH 12.7~$\micron$/PAH 11.3~$\micron$ ratios depend more strongly on the PAH edge structure rather than the degree of PAH ionization in M17.

Comparing Figs.~\ref{othermap2}(e)-\ref{othermap2}(h) in detail, we find that there are small differences in the relation; the PAH 12.0~$\micron$/PAH 11.3~$\micron$ and PAH 13.5~$\micron$/PAH 11.3~$\micron$ ratios increase nearly in proportion to the PAH 12.7~$\micron$/PAH 11.3~$\micron$ and PAH 14.2~$\micron$/PAH 11.3~$\micron$ ratios, respectively (Figs.~\ref{othermap2}(e) and \ref{othermap2}(h)).
On the other hand, the PAH 13.5~$\micron$/PAH 11.3~$\micron$ and PAH 14.2~$\micron$/PAH 11.3~$\micron$ ratios do not increase in proportion to the PAH 12.7~$\micron$/PAH 11.3~$\micron$ ratio, and there are apparent offsets in their relations (Figs.~\ref{othermap2}(f) and \ref{othermap2}(g)).
Hence, the PAH 12.0~$\micron$ and 12.7~$\micron$ intensities start to increase first and then increases in the PAH 13.5~$\micron$ and 14.2~$\micron$ intensities follow, relative to the PAH 11.3~$\micron$ intensity.
The relative delay in the growth of the PAH 13.5~$\micron$ and 14.2~$\micron$ intensities suggests that processing of PAHs may gradually take place due to the intense (and/or hard) UV.
The proportionality of the PAH 13.5~$\micron$ intensity to the PAH 14.2~$\micron$ intensity (Fig.~\ref{othermap2}(h)) supports that they are of the same origin (quartet C-H in-plane bending mode), as suggested by \citet{Tielens05}.

\section{Conclusion}

Based on $Spitzer$/IRS spectral mapping observations, we have examined spatial variations of the PAH properties around the M17SW region.
We analyzed independent 990 spectra, which show prominent PAH features at wavelengths of 6.2, 7.7, 8.6, 11.3, 12.0, 12.7, 13.5, and 14.2~$\micron$ as well as fine-structure lines.
We decomposed all the spectral features using PAHFIT (\citealt{Smith07a}).
As a result, the derived PAH emissions are bright in regions between HII regions traced by Br$\gamma$ and molecular cloud regions traced by $^{13}$CO.
Additionally, the PAH intensity maps show a large-scale gradient from the north-east to south-west, indicating that PAHs are irradiated by UV from the OB stars in the M17 center.
By contrast, PAH interband ratio maps show no clear large-scale gradient but they are locally changed.
These results suggest that the PAH ionization is mainly controlled by local conditions rather than the large-scale UV environment determined by the OB stars in the M17 center.
The degree of PAH ionization estimated from the PAH 7.7~$\micron$/PAH 11.3~$\micron$ ratios ranges from 19 to 81~\% with a median value of 48~\%, which is comparable to that in the previous studies for NGC~7023 and the Orion Bar.
We also find that the degree of PAH ionization is high near the peak of the $^{13}$CO emission.
We discuss the ionization balance using the $G_0$ and [CII] maps to find that $G_0$/[CII] ratios show a local maximum inside the molecular cloud.
We conclude that buried B-type or later stars may be important to determine the degree of PAH ionization in local conditions.
Additionally, the PAH edge structures are examined by ratios of the PAH~12.0, 12.7, 13.5, and 14.2~$\micron$ features to the PAH~11.3~$\micron$ feature, which suggests that edgy PAHs are processed due to the intense (and/or hard) UV radiation especially near the M17 center.

\acknowledgments

This work is based on archival data obtained with the $Spitzer$ Space Telescope, which is operated by the Jet Propulsion Laboratory, California Institute of Technology under a contract with NASA, and with the $Herschel$ Space Observatory.
This work is also based on observations with IRSF and the 45-m telescope in the Nobeyama Radio Observatory (NRO).
The operation of IRSF is supported by Joint Development Research of National Astronomical Observatory of Japan, and Optical Near-Infrared Astronomy Inter-University Cooperation Program, funded by the the Ministry of Education, Culture, Sports, Science and Technology of Japan.
NRO is a branch of the National Astronomical Observatory of Japan, National Institutes of Natural Sciences.
This research was supported by JSPS KAKENHI Grant Number 25247020.
S.~Ohashi is financially supported by a Research Fellowship from JSPS for Young Scientists.
We also express many thanks to the anonymous referee for the useful comments.

\end{document}